\documentstyle[12pt,epsfig]{article}
\newcommand{\be}{\begin{equation}}

\newcommand{\dms}{\mbox{$\Delta m^2_{\odot}$}}
\newcommand{\dma}{\mbox{$\Delta m^2_{\rm A}$}}
\newcommand{\en}{\end{equation}}
\newcommand{\beeq}{\begin{eqnarray}}
\newcommand{\eneq}{\end{eqnarray}}
\newcommand{\eps}{\varepsilon}

\makeatletter
\def\vereq#1#2{\lower3pt\vbox{\baselineskip1.5pt \lineskip1.5pt
\ialign{$\m@th#1\hfill##\hfil$\crcr#2\crcr\sim\crcr}}}
\makeatother
\begin{document}
\begin{titlepage}
\begin{flushright}
TUM-HEP-622/06
\end{flushright}
\begin{center}
{\Large\bf 
5D seesaw, flavor structure, and mass textures
} 
\end{center}
\vspace{1cm}
\begin{center}
Naoyuki {Haba}
\footnote{E-mail: haba@ph.tum.de \\
On leave of absence from 
 Institute of Theoretical Physics, University of
 Tokushima, 770-8502, Japan}
\end{center}
\vspace{0.2cm}
\begin{center}
{\it Physik-Department, 
 Technische Universit$\ddot{a}$t M$\ddot{u}$nchen,
 James-Franck-Strasse,
 D-85748 Garching, Germany}
\\[0.2cm]
\end{center}
\vspace{1cm}
\begin{abstract}

In the 5D theory in which 
 only 3 generation right-handed neutrinos
 are in the bulk, 
 the neutrino flavor mixings and the mass spectrum
 can be constructed through the 
 seesaw mechanism. 
The 5D seesaw 
 is easily calculated just by a
 replacement of the 
 Majorana mass eigenvalues, $M_i$, 
 by $2 M_* \tan({\rm h})[\pi RM_i]$ ($M_*$: 5D Planck scale,
 $R$: compactification radius). 
The 5D features
 appear 
 when
 the bulk mass, which induces
 the 4D Majorana mass, 
 is the same as the compactification scale 
 or
 larger than it.
Depending on the 
 type of bulk mass, 
 the seesaw scales of the 3 generations 
 are
 strongly split (the tan-function case) or 
 degenerate (the tanh-function case).
In the split case,  
 the seesaw enhancement is
 naturally realized. 
The single right-handed neutrino
 dominance works in a simple
 setup, and 
 some specific mass textures,
 which are just assumptions in the 4D setup,
 can be naturally obtained in 
 5 dimensions. 
The 
 degenerate case 
 is also useful for a suitable 
 neutrino flavor structure.

\end{abstract}
\end{titlepage}
\newpage
\section{Introduction}

Recently, 
 the neutrino oscillation experiments
 show the detailed flavor structure in the
 lepton sector\cite{Maltoni:2004ei}\cite{Maltoni:2004ei2}. 
This is one of the most
 important keys to find 
 physics 
 beyond the standard model (SM). 
The smallness of the neutrino mass itself
 is also a clue for 
 new physics.
The seesaw mechanism with the
 heavy Majorana masses of right-handed
 neutrinos 
 is one of the
 most reliable candidates
 for the underlying theory of the SM\cite{seesaw}. 
The large flavor mixings
 in the lepton sector
 can also be realized by 
 the seesaw mechanism. 
For example,
 the large Majorana mass hierarchy
 can induce the large flavor mixings
 through the seesaw mechanism,
 even when the Dirac and Majorana neutrino 
 mass matrices have small flavor
 mixings\cite{enhance}.
This is the typical
 example of a so-called 
 seesaw enhancement\cite{enhance2}. 
The single right-handed neutrino
 dominance (SRND)\cite{King:1999cm}
 should be classified as the seesaw
 enhancement,
 in which the lightest Majorana mass is 
 the dominant contribution of the
 seesaw and induces the large flavor
 mixing with the Dirac mass matrix structure. 
There are a lot of other attempts to
 obtain 
 the suitable neutrino flavor structure
 through the seesaw mechanism from
 the viewpoint of 
 mass 
 textures, flavor symmetries,
 and so on.

On the other hand,
 the large extra dimensional
 theory has been proposed to solve
 the gauge hierarchy problem\cite{Arkani-Hamed:1998rs}.
This theory
 explains the smallness of 
 Dirac neutrino mass due to  
 the existence of right-handed 
 neutrinos in 
 the bulk\cite{Arkani-Hamed:1998vp0}. 
It suggests 
 new possibilities for
 the neutrino
 physics\cite{Arkani-Hamed:1998vp}.
In the framework of the extra
 dimensional theory,
 we can also consider the seesaw mechanism 
 by introducing a bulk mass for the
 right-handed neutrinos
 which induces the 
 4D Majorana 
 mass\cite{Dienes:1998sb}-\cite{Kim:2003vr}.
%

In this paper,
 we adopt the 
 5D setup\footnote{The neutrino flavor structure
 has been discussed in the similar setup by
 introducing the $S_3$ flavor
 symmetry and its breaking through a boundary 
 twist\cite{tri5d}. } 
 in which the 
 3 generation 
 right-handed neutrinos
 are in the bulk 
 with bulk mass.
We do not specify the size of 
 the compactification scale
 so that our discussions 
 are applicable to both 
 large and small
 extra dimensional theories. 
We show 
 a simple calculation method 
 of the 5D seesaw, 
 which just replaces 
 the 4D Majorana mass eigenvalues, $M_i$, 
 by $2M_* \tan({\rm h})[\pi RM_i]$ ($M_*$: 5D Planck scale,
 $R$: compactification radius). 
The neutrino flavor mixings and the mass spectrum
 can be drastically 
 changed by the 
 5D seesaw. 
The 5D features
 appear 
 when
 the Majorana mass is the same
 as the compactification scale 
 or
 larger than it.  
Depending on the 
 type of bulk mass, 
 the seesaw scales of the 3 generations  
 are strongly split (the tan-function case) or 
 degenerate (the tanh-function case).
In the split case, 
 the seesaw enhancement
 is realized naturally. 
We show that 
 the SRND 
 works in a simple
 setup in 5 dimensions. 
This split also 
 makes 
 the tiny symmetry breaking 
 induce the suitable flavor structure
 in the $L_\mu-L_\tau$
 symmetry\cite{lmlt_o}\cite{Rodejohann:2005ru}. 
We also show that 
 some specific mass textures,
 which the 4D theory 
 just assumes,  
 can be naturally obtained 
 in the 5D seesaw. 
The 
 degenerate case  
 can also 
 realize the suitable 
 flavor structure.

\section{Setup and 5D seesaw}

We take the 5D setup, 
 where the 5th dimension coordinate $(y)$
 is compactified on $S^1/Z_2$.
Only 
 right-handed neutrinos,
 which are the SM gauge singlets,
 are spread in the 5D bulk, while
 the other quarks and leptons are
 localized on the 4D brane.
Since the 5D theory is a vector-like
 theory, we must introduce 
 the chiral partners, $N^c_i$, of the 
 right-handed neutrinos, $N_i$,
 where $N^c_i$ and $N_i$ are 
 the two component Weyl spinors of the $i$th generation 
 ($i=1, 2, 3$). 
Under $Z_2$ parity, $y \rightarrow -y$,
 ${\cal N}\equiv(N,N^c)^T$
 transforms as
 ${\cal N}(x^\mu, -y)= \gamma^5 {\cal N}(x^\mu, y)$.
Then the mode expansion is given by
\begin{eqnarray}
&& {\cal N}_i(x^\mu,y) \;=\; \frac{1}{\sqrt{\pi R}} \left(
  \begin{array}{r}
    \frac{1}{\sqrt{2}}\,N_i^{(0)}(x^\mu)+
    \sum\limits_{n=1}^\infty
    \cos (\frac{ny}{R})\,N_i^{(n)}(x^\mu) \\
    \sum\limits_{n=1}^\infty
    \sin (\frac{ny}{R})\,N^c_i{}^{(n)}(x^\mu)
  \end{array} \right) ,
\label{mode11}
\end{eqnarray}
where the factor $1/\sqrt{2}$ of the zero-mode
 is needed for the canonical
 kinetic term in the 4D effective Lagrangian.
Hence,
 only right-handed neutrinos, $N_i$'s, have
 the zero-modes and survive below the
 compactification scale, $R^{-1}$.

The general 
 5D bulk mass terms which induce the 4D Majorana masses 
 are given by\cite{Lukas:2000rg} 
\begin{equation}
\label{3.8}
  {\cal L}_{\rm 5DM} \;=\;
  -\frac{1}{2}\left(M^S_{ij} \; \overline{{\cal N}^c_i}{\cal N}_j
  +M^V_{ij} \; \overline{{\cal N}^c_i}\gamma_5{\cal N}_j
 +{\rm h.c.}\right) .
\end{equation}
Here, 
 $M^S$ and $M^V$ are symmetric matrices for
 the generation index, $i,j$.
They are invariant under the 
 $Z_2$ parity but
 break the lepton number. 
$M^V$ is not 
 5D Lorentz invariant but 
 derivable from 
 the spontaneous symmetry breaking 
 of the 5D Lorentz invariance\cite{Lukas:2000rg}.

By integrating out the 5th dimension, 
 the 
 4D effective Lagrangian is obtained, in which 
 the Majorana mass terms  
 are given by
\begin{equation}
  {\cal L}_{\rm 4DM}\, \;=\; 
  -{1\over2}\left(
 \sum\limits_{n=0}^\infty M^A_{ij}
            {N_i}^{(n)T}{N_j}^{(n)}+
 \sum\limits_{n=1}^\infty M^{B*}_{ij}
            \overline{N_i^c}^{(n)T}\overline{N_j^c}^{(n)}
          \right) +{\rm h.c.},
\end{equation}
where $M^A\equiv M^S+M^V$ and $M^{B*}\equiv -M^{S*}+M^{V*}$. 
We omit spinor indices here. 
Kaluza-Klein (KK) masses are given by
\begin{eqnarray}
&&  {\cal L}_{\rm KK} \;=\; -\sum\limits_{n=1}^\infty
  \; \frac{n}{R}\;(\overline{N_i^c}^{(n)}{N_i}^{(n)}
  +\overline{N_i}^{(n)}{N_i^c}^{(n)}). 
\end{eqnarray}
The 
 Dirac mass terms between the bulk
 right-handed neutrinos, $N_i$ and $N_i^c$, and
 the brane-localized lepton doublets, $L_i$,
 are given by
\begin{equation}
{\cal L}_{\rm 4Dm}=
-\frac{1}{\sqrt{M_*}}(\overline{L_i}\:m_{ij}\:N_j\:
 +\:\overline{L_i}\:m_{ij}^c\:N_j^c\:)\delta(y)+{\rm h.c.}.
\label{4Dm}
\end{equation}
In our setup, $m_{ij}^c=0$ due
 to the $Z_2$ parity
 (mode expansion in Eq.(\ref{mode11})).
Then the 4D neutrino mass
 matrix 
 is given by 
\begin{eqnarray}
 {\cal L}_{m_\nu} \; & = \; & -\frac{1}{2}\, \nu_i^T\,
  ({\cal M}_\nu)_{ij}\, \nu_j \,+{\rm h.c.} , \nonumber \\
  {\cal M}_\nu
&=&
\left(\begin{array}{ccccc|cccc}
  0 & 0& 0 &    \cdots& & 
 \frac{m_{ij}}{\sqrt{2\pi RM_*}} & 
 \frac{m_{ij}}{\sqrt{\pi RM_*}} & \frac{m_{ij}}{\sqrt{\pi RM_*}} 
 & \cdots   \\
    & M^{B*}_{ij} &  &  & &  &\frac{1}{R} & &  \\
    &  & M^{B*}_{ij} &  & &  & &\frac{2}{R} &   \\
   \vdots & & &  \ddots & & & &  &\ddots \\ \hline
   \frac{(m^T)_{ij}}{\sqrt{2\pi RM_*}} &  &   & & & M^{A}_{ij} & & &   \\
  \frac{(m^T)_{ij}}{\sqrt{\pi RM_*}}
   & \frac{1}{R} &  & &  & & M^{A}_{ij} & &   \\
   \frac{(m^T)_{ij}}{\sqrt{\pi RM_*}} 
   &  & \frac{2}{R} & &  &  & & M^{A}_{ij} &   \\
   \vdots &  &  & \ddots & &  & &  &\ddots  
\end{array} \right),
\label{7}
\end{eqnarray}
in the basis of
$\nu_i^T\equiv ( \overline{L_i}, \overline{N^{c}_i}^{(1)},
 \overline{N^{c}_i}^{(2)},\cdots |
 N^{(0)}_i, N^{(1)}_i, N^{(2)}_i, \cdots )$. 
Let us adopt $M^V$ 
 for the bulk mass, and
 pick up the sub-matrix of $n$-mode from Eq.(\ref{7}). 
Since the KK mass, $n/R$, is proportional to 
 the unit matrix in the flavor space,
 all $n$-modes of the 
 right-handed neutrinos are 
 diagonalized simultaneously 
 in the flavor space as
\begin{equation}
\left(
\begin{array}{cc}
U_R^\dagger &  \\
 & U_R^T       
\end{array}
\right)  \;
\left(
\begin{array}{cc}
M^{V*}_{ij} & \frac{n}{R} \\
\frac{n}{R} & M^V_{ij}     
\end{array}
\right)  \;
\left(
\begin{array}{cc}
U_R^* &  \\
 & U_R       
\end{array}
\right) ,
\label{MV22}
\end{equation}
where 
 $U_R$ is the unitary matrix 
 which
 diagonalizes $M^V_{ij}$ as 
 $U_R^T M^V_{ij}U_R = M_d$.  
The three eigenvalues of $M_d$ 
 can be written as $M_i = |M_i| e^{i\alpha_i}$ 
 $(i=1,2,3)$\footnote{
We take the basis where
 the phase $\alpha_i$ 
 remains in $M_i$ 
 in Eqs.(\ref{inverse}) and (\ref{9}).
It is also possible to absorb this phase
 by using the unitary matrix 
 $U_R P$, 
 where $P_{ii}=e^{-i\alpha_i/2}$. 
In this basis 
 $M_i$ is real and positive in Eq.(\ref{MV22}), and 
 then 
 the Dirac mass matrix becomes 
 $m_R P$ which induces 
 the phase $e^{-i\alpha_k}$
 in Eq.(\ref{9}).}. 
In this basis, the Dirac mass matrix
 is given by $(0\;\; m U_R)$. 
We denote 
 $m_R \equiv m U_R$
 hereafter. 
Since the inverse mass matrix of Eq.(\ref{MV22})
 is given by  
\begin{equation}
{1 \over |M_i|^2-(\frac{n}{R})^2}
\left(
\begin{array}{cc}
M_i & -\frac{n}{R} \\
-\frac{n}{R} & M_i^* 
\end{array}
\right) , 
\label{inverse}
\end{equation}
the summation of the infinite numbers
 of ``seesaw'' is calculated as 
\begin{eqnarray}
\label{9}
m^{\nu}_{ij}&=&
-\frac{(m_R){}_{ik}}{\pi RM_*}\:  \left[ 
\frac{1}{2M_k}+
\sum_{n=1}^{\infty}
{M_k^* \over |M_k|^2-(\frac{n}{R})^2}
\right]
 \: (m_R^T)_{kj} \nonumber \\
&=& -(m_R)_{ik}\:
 {e^{-i\alpha_k} \over 2 M_* \tan[\pi R|M_k|]}\:(m_R^T)_{kj}, 
\end{eqnarray}
when the magnitude of  
 ${\rm min}[|M_i| \pm \frac{n}{R}]$
 is much 
 larger than the Dirac mass scale. 
The case of 
 $|M_i| \ll R^{-1}$  
 reproduces 
 the ordinal 4D seesaw formula with 
 the volume suppression factor,
 ${2\pi RM_*}$.

To give a short summary, 
 in order to get the 5D seesaw mass matrix, 
 we just make the replacement 
\begin{equation}
\label{seesaw5D3}
{M_i} 
 \rightarrow 
 {2 M_* \tan [\pi R|M_i|]}\; e^{i \alpha_i}   
\end{equation}
in the Majorana mass diagonal basis. 
By use of Eq.(\ref{seesaw5D3}),
 we can immediately 
 achieve the 
 5D seesaw calculations. 
When we adopt $M^S$ for a bulk mass in 
 Eq.(\ref{3.8}),
 the replacement in Eq.(\ref{seesaw5D3}) 
 is modified as 
\begin{equation}
\label{11}
{M_i} 
 \rightarrow 
 {2 M_* \tanh [\pi R|M^S_i|]}\; e^{i \beta_i}, 
\end{equation}
where $M^S_i (=|M^S_i|e^{i\beta_i})$
 is the mass eigenvalue of $M^S_{ij}$\footnote{ 
 Equation (\ref{seesaw5D3}) ((\ref{11})) is 
 consistent with the results in 
 Ref.\cite{Dienes:1998sb} (\cite{Lukas:2000rg}), 
 where $M^{V}$ ($M^S$) is
 taken to be real.}. 
Anyhow, the tan(h)-function is 
 a feature in the 5D seesaw,
 which has the significant effects
 when $|M_i|, |M^S_i| \geq {\cal O}(R^{-1})$.
The typical case is
 $|M_i|=|\frac{2n+1}{2R}|$
 ($n$: integer) 
 where
 the light 
 neutrino becomes massless. 
It is because 
 the infinite times seesaw becomes
$-\frac{1}{2\pi R M_*} \times$
$({[\frac{1}{2R}]^{-1}}-{[\frac{1}{2R}]^{-1}}
+{[\frac{3}{2R}]^{-1}}-{[\frac{3}{2R}]^{-1}}
+\cdots ) \rightarrow 0$.
Notice that 
 $n$ should satisfy 
 $|n+\frac{1}{2}|\ll RM_*$ 
 to fulfill $|M_i| \ll M_*$.

When $|M_i| \geq {\cal O}(R^{-1})$,
 the largest contribution to the seesaw 
 is not from $|M_i|$ but from 
 ${\rm min}[|M_i| \pm \frac{n}{R}] (\leq \frac{1}{2R})$, 
 which 
 is the origin of 
 the periodic tan-function\cite{Dienes:1998sb}. 
On the other hand, 
 the 5D Planck scale 
 plays a role of 
 seesaw scale when 
 $|M_i^S| \geq {\cal O}(R^{-1})$.
Therefore, 
 in the case of $|M_i|, |M_i^S| \geq {\cal O}(R^{-1})$,
 the hierarchy of the 
 seesaw scales 
 can be different from 
 the original mass hierarchies 
 of $M_i$, $M^S_i$.
For example, suppose 
 $M_1^{(S)} < M_2^{(S)} < M_3^{(S)}$
 for the 
 original Majorana mass hierarchy,
 then, the 
 hierarchy of the 5D seesaw scales 
 becomes acceleratingly split,
 or even arbitrary from the
 periodicity, (degenerate) due to 
 the tan(h)-function. 

If we adopt both $M^V$ and $M^S$
 for the bulk mass simultaneously, 
 the seesaw calculation becomes
 complicated. 
However, if they are 
 diagonalized by the
 same unitary matrix $U_R$, 
 we can obtain the simple
 5D seesaw formula. 
Taking the Majorana mass eigenvalues
 to be real for simplicity\footnote{
If they are complex,
 the formula is given by
 $-(m_R)_{ik}\frac{(M_k^*-M^{S*}_k)}{2M_*M'_k \tan[\pi RM'_k]}(m_R^T)_{kj}$ 
 where $M'_k \equiv ((M_k^*-M_k^{S*})(M_k+M^S_k))^{1/2}$.
 We must take the same phase for two ${M'_k}$s 
 (square root of the complex number) in the
 denominator.},
 the formula becomes 
\begin{eqnarray}
&&m^{\nu}_{ij}
= -(m_R)_{ik}\:
 {M_k-M_k^{S} \over 2 M_* M_c
  \tan[\pi RM_c]}\:(m_R^T)_{kj}
 \;\;\;\;\;\;\;\;\;
 (|M_k|>|M^{S}_k|), 
\label{mkms} \\ 
&&\;\;\;\;\;\;= -(m_R)_{ik}\:
 {M_k^{S}-M_k \over 2 M_* M_c'
  \tanh[\pi RM_c']}\:(m_R^T)_{kj}
 \;\;\;\;\;\;\;\: (|M_k|<|M^{S}_k|),
\label{msmk} 
\end{eqnarray}
where $M_c \equiv \sqrt{M_k^2-M_k^{S2}}$ and
 $M_c'\equiv \sqrt{M_k^{S2}-M_k^2}$. 
When 
 $|M_k|>|M_k^S|$ ($|M_k|<|M_k^S|$),
 the denominator has a tan(h)-function. 
Thus, it is possible that
 different generations have 
 different seesaw scales characterized by
 the tan- or tanh-function in general. 
We take 
 either 
 $M^V$ or $M^S$
 for a bulk mass
 of the right-handed neutrinos
 in the following discussions,  
 but we can always replace $M^V$ ($M^S$)
 by two real masses of $M^V$ and $M^S$ with Eq.(\ref{mkms}) 
 (Eq.(\ref{msmk})).

\section{Flavor structure from 5D seesaw}

In this section we 
 show some 5D seesaw examples 
 which are just assumptions
 or impossible in the 4D setup. 
We show useful examples of the 
 tan(h)-function for the neutrino
 flavor structure.


\subsection{Mass textures}

At first, we show two examples
 of neutrino mass textures when 
 one Majorana mass eigenvalue 
 satisfies 
 $|M_i|=|\frac{2n+1}{2R}|$\footnote{
 The Majorana mass is not necessarily exact  
 $|M_i| =|\frac{2n+1}{2R}|$.  
Close enough to 
 this value the following mechanisms
 work well.} ($|n+\frac{1}{2}|\ll RM_*$). 
This case makes the $i$th column 
 of the Dirac mass matrix, $m_{ij}$, 
 vanish from the seesaw calculation.

For a specific mass texture,
 there should exist the 
 flavor symmetry
 as the symmetry of the underlying theory of the
 SM.
Here, 
 let us consider the $S_3$ flavor 
 symmetry\cite{S30} \cite{S3}.
The setup is designed in a way  
 that both right- and 
 left-handed neutrinos are
 ${\bf 3}(={\bf 2}+{\bf 1_S})$ 
 representations
 of $S_3$,
 in which 
 the 1st and 2nd generations are 
 doublets (${\bf 2}$) and the 3rd generation is 
 the singlet (${\bf 1_S}$) in the complex 
 basis. 
We consider two kinds of
 ${\bf 3}(={\bf 2}+{\bf 1_S})$ 
 $S_3$-Higgs fields, $\phi^D$ and $\phi^N$, in the bulk, 
 where we denote the doublet as $\phi^{D,N}_{1,2}$ and 
 the singlet as $\phi^{D,N}_{S}$. 
$\phi^N$ possesses the lepton number while 
 $\phi^D$ does not. 
We set $\phi^D$ as a $SU(2)_L$ singlet 
 to avoid large lepton flavor
 violations (and also to use $\phi^*$ in
 Eq.(\ref{mdirac})\footnote{
When $\phi$ has some other quantum charges 
 than $S_3$,
 $\phi^*$ is removed from
 the mass matrix elements\cite{S3}.}). 
Then, 
 the general Dirac and Majorana mass matrices
  become\cite{S3}
\begin{eqnarray}
  m &=& \left(
\begin{array}{ccc}
  x \phi^D_S+ x' \phi^{D*}_S 
 & z\phi^D_1+z'\phi^{D*}_2 & w\phi^D_2+\xi'\phi^{D*}_1\, \\
  z\phi^D_2+z'\phi^{D*}_1 & x \phi^D_S+x' \phi^{D*}_S  &
   w\phi^D_1+\xi'\phi^{D*}_2\, \\ 
  \xi\phi^D_1+w'\phi^{D*}_2 & \xi\phi^D_2+w'\phi^{D*}_1
    & y \phi^D_S+y' \phi^{D*}_S 
\end{array}
\right),
\label{mdirac} \\
  M &=& 
\left(
\begin{array}{ccc}
  \,\alpha \phi^N_1 & 
     \gamma\phi^N_S & \eta\phi^N_2\, \\
  \,\gamma\phi^N_S & \alpha\phi^N_2 & 
                 \eta\phi^N_1\, \\
  \,\eta\phi^N_2 & \eta\phi^N_1 & \rho\phi^N_S \,
  \end{array}
   \right),
  \label{Mmajorana}
\end{eqnarray}
respectively. 
Where 
$\phi^D$ and $\phi^N$
 stand for their vacuum expectation
 values (VEVs) of zero-modes,
 which could be determined by
 the $S_3$-Higgs potential 
 analysis in principle. 
$x, x', 
 \cdots, \eta^{},\rho^{}$
 are independent Yukawa couplings, 
 and those of Eq.(\ref{mdirac}) ((\ref{Mmajorana}))
 are proportional to 
 $\frac{\langle H \rangle}{2\pi RM_*^2}$
 ($\frac{1}{\sqrt{2\pi RM_*}}$), 
 where 
 $H$ is the SM Higgs field. 
For the Dirac mass matrix,
 $L_i$ ($N_i$) is taken from
 the left- (right-) hand side
 as in Eq.(\ref{4Dm}).

\vspace{2mm}

The first example is 
 the neutrino mass texture which induces
 the inverted hierarchy (IH) mass 
 spectrum\cite{Jezabek:1998du}.
We introduce only $S_3$ singlet 
 Higgs, $\phi^N_S$, for the
 Majorana mass.
As for the Dirac sector, 
 we introduce both, $\phi^D_S$ and $\phi^D_{1,2}$,
 and 
 assume $z=w'=0$
 as well as a vanishing VEV, 
  $\phi^D_1 =0$.
Then
 the Dirac and Majorana mass matrices are
 given by 
\begin{equation}
m = 
\left(
\begin{array}{ccc}
A & E & D \\
0 & B & F\\
0 & C & G
\end{array}
\right),  \;\;\;
M = \left(
\begin{array}{ccc}
 & X & \\
X &  &  \\
 &  & Y
\end{array}
\right) ,
\end{equation}
respectively. 
We denote 
 $X=|X|e^{i\theta_X}$ and 
 $Y=|Y|e^{i\theta_Y}$. 
In the diagonal basis of
 Majorana mass matrix, the matrices are
 rewritten as
\begin{equation}
m_R = \frac{1}{\sqrt{2}}
\left(
\begin{array}{ccc}
A\hspace{-1mm}-\hspace{-1mm}E & A\hspace{-1mm}+\hspace{-1mm}E & \sqrt{2}D \\
-B & B & \sqrt{2}F\\
-C & C & \sqrt{2}G
\end{array}
\right),  \;\;\;
M_d = \left(
\begin{array}{ccc}
-\hspace{-1mm}X &  & \\
 & X &  \\
 &  & Y
\end{array}
\right) ,
\end{equation}
which 
 suggest for the 5D seesaw, 
\begin{eqnarray}
m_\nu^{\rm 5D}
&=& - m_R \: \left(M_d\right)^{-1} \: 
  m_R^T, \nonumber \\
&=& -\frac{1}{4M_*}
\left[
\delta_1
\left( 
\begin{array}{c}
A\hspace{-1mm}-\hspace{-1mm}E  \\ 
-B  \\ 
-C  
\end{array} 
\right)
(A\hspace{-1mm}-\hspace{-1mm}E\; -\hspace{-1mm}B\; -\hspace{-1mm}C) +
\delta_2
\left( 
\begin{array}{c}
A\hspace{-1mm}+\hspace{-1mm}E  \\ 
B \\ 
C 
\end{array} 
\right)
(A\hspace{-1mm}+\hspace{-1mm}E\; B \; G ) 
\right. \nonumber \\
&& \;\;\;\;\;\;\;\;\;\;\;\;\;\;\;\;\;\;\;\;\;\;\;\;
\;\;\;\;\;\;\;\;\;\;\;\;\left. +
\delta_3
\left( 
\begin{array}{c}
\sqrt{2}D  \\ 
\sqrt{2}F \\ 
\sqrt{2}G 
\end{array} 
\right)
( \sqrt{2}D\; \sqrt{2}F\; \sqrt{2}G )
\right] , 
\label{sd516}
\end{eqnarray}
where
 $\delta_1 = -\delta_2 =-\cot[\pi R |X|] e^{-i\theta_X}$ and
 $\delta_3 = \cot[\pi R |Y|]e^{-i\theta_Y}$.
The $i$th term 
 is the seesaw 
 contribution from $M_i$ ($i=1,2,3$). 
In the case of 
 $|Y|=|\frac{2n+1}{2R}|$\footnote{
Using $\langle \phi^N_S \rangle$,
 this condition is rewritten as
 $|\rho \langle \phi^N_S \rangle |
 =\sqrt{\frac{\pi(2n+1)^2}{2}R^{-1}M_*}$.
},
 the 3rd generation right-handed
 neutrino becomes heavy and decouples, and hence 
 the 3rd term of Eq.(\ref{sd516}) vanishes.
Then, the neutrino mass texture
 becomes
\begin{equation}
m_\nu^{\rm 5D}\simeq 
\frac{\delta_1}{2M_*}
\left(
\begin{array}{ccc}
2AE & AB & AC \\
AB & 0 & 0 \\
AC & 0  &  0
\end{array}
\right) .
\end{equation}
The suitable mass texture
 for the IH mass spectrum 
 is obtained 
 when $|E| \ll |A|, |B|, |C|$ 
 and 
 $|B| \simeq |C|$. 

\vspace{2mm}

The second example is 
 the neutrino mass texture with 
 only two right-handed neutrinos\cite{Frampton:2002qc}.
The introduction of only two 
 right-handed neutrinos
 seems artificial.
However, in the 
 5D setup, 
 this texture can be 
 naturally realized as follows:

Introducing only an $S_3$ doublet Higgs, $\phi^D_{1,2}$, 
 the diagonal elements of the Dirac mass matrix 
 vanish. 
As for the Majorana mass matrix,
 we introduce both $\phi^N_S$ and $\phi^N_{1,2}$,
 and assume 
 $\gamma =\eta =0$.
Then the Dirac and Majorana mass matrices are
 given by 
\begin{equation}
m = 
\left(
\begin{array}{ccc}
0 & A & B \\
C & 0 & D\\
E &  F  & 0
\end{array}
\right),  \;\;\;
M = \left(
\begin{array}{ccc}
X & & \\
 & X'&  \\
 &  & Y
\end{array}
\right) ,
\end{equation}
respectively, 
 where we denote 
 $X=|X|e^{i\theta_X}$, 
 $X'=|X'|e^{i\theta_X'}$, and 
 $Y=|Y|e^{i\theta_Y}$. 
Then, the 5D seesaw is given by 
\begin{eqnarray}
&&\hspace*{-3mm}m_\nu^{\rm 5D}\hspace{-1mm}=\hspace{-1mm} 
-\frac{1}{2M_*}\hspace{-1mm}
\left[
\delta_1'
\left( 
\begin{array}{c}
0  \\ 
C  \\ 
E  
\end{array} 
\right)
(0\; C\; E) +
\delta_2'
\left( 
\begin{array}{c}
A  \\ 
0 \\ 
F 
\end{array} 
\right)
(A\; 0 \; F ) 
+
\delta_3'
\left( 
\begin{array}{c}
B  \\ 
D \\ 
0 
\end{array} 
\right)
(B\; D\; 0 )
\right]\hspace{-1mm},
\label{sd5}
\end{eqnarray}
where
 $\delta_1' = {\cot[\pi R |X|]}e^{-i\theta_X}$, 
 $\delta_2' = {\cot[\pi R |X'|]}e^{-i\theta_X'}$, and 
 $\delta_3' = {\cot[\pi R |Y|]}e^{-i\theta_Y}$.
In the case of 
 $|X'| =|\frac{2n+1}{2R}|$,  
 the 2nd generation right-handed neutrino
 decouples and 
 the 2nd term in Eq.(\ref{sd5}) 
 vanishes. 
Then the neutrino mass texture
 becomes 
\begin{equation}
m_\nu^{\rm 5D}= 
-\frac{\delta_1'}{2M_*}
\left(
\begin{array}{ccc}
r B^2 & r BD & 0 \\
r BD & r D^2\hspace{-1mm} +\hspace{-1mm} C^2 &  CE \\
0 &  CE  &      E^2 
\end{array}
\right) ,
\end{equation}
where $r \equiv \frac{\delta_3'}{\delta_1'}$.
For $|B|\sim |D|$ and $|C| \simeq |E|$ with 
 $|r| \sim 0.1 \times \frac{|C^2|}{|D^2|}$, 
 bi-large mixing with
 the normal hierarchy (NH) mass spectrum 
 is realized\cite{Frampton:2002qc}.


\subsection{SRND in $L_\mu - L_\tau$ symmetry}

When $|M_i| \sim |\frac{2n+1}{2R}|$, 
 the tan-function 
 strongly splits 
 the 
 effective seesaw scales, which 
 is useful for the
 SRND. 
As an example of the SRND,
 let us discuss 
 the neutrino 
 mass texture in the 
 $L_\mu - L_\tau$ symmetry.
In the 4D case,
 the seesaw mechanism needs a large
 symmetry breaking parameter in
 the Majorana mass matrix to realize
 the NH mass spectrum\cite{Rodejohann:2005ru}. 
On the other hand, 
 the 5D seesaw
 can enhance 
 a small symmetry breaking 
 and realize 
 the NH mass spectrum with the suitable 
 flavor mixings as follows:

At first, 
 let us show 
 the 2-3 generation sub-matrix. 
Under the $L_\mu - L_\tau$ symmetry, 
 the 
 Dirac and Majorana mass
 matrices are given by\cite{Rodejohann:2005ru}  
\begin{eqnarray}
m & = & 
\left(
\begin{array}{cc}
A & 0 \\
0 & B \\
\end{array} 
\right), \;\;\;
M =
\left(
\begin{array}{cc}
\eps & X \\
X & \eps \\
\end{array} 
\right),
\end{eqnarray}
respectively,  
 where $A,B,X$ are
 $L_\mu - L_\tau$ symmetric 
 mass parameters.
We take $X$ to be 
 real and positive
 for simplicity.
$\eps \:(>0)$ is the symmetry
 breaking parameter
 which 
 should be smaller
 than the symmetric one $X$.
In the basis where
 the Majorana mass matrix is diagonal 
 the seesaw formula is given by 
\begin{eqnarray}
m_{\nu}  =  
-\frac{1}{2}
\left(
\begin{array}{cc}
A & A \\
-B & B \\
\end{array} 
\right) 
\left(
\begin{array}{cc}
{1\over \eps -X} & 0 \\
0 & {1\over \eps +X} \\
\end{array} 
\right) 
\left(
\begin{array}{cc}
A & -B \\
A &  B \\
\end{array} 
\right) .
\label{222}
\end{eqnarray}
For the realization of a NH mass spectrum,
 large 
 symmetry
 breaking, $\eps \sim X$, 
 is required\cite{Rodejohann:2005ru}.

The 
 5D seesaw 
 can avoid 
 this unnatural large symmetry breaking. 
By using Eq.(\ref{seesaw5D3}), 
 we can obtain the 
 5D seesaw directly from 
 Eq.(\ref{222}) as 
\begin{eqnarray}
m_{\nu}^{\rm 5D}  =  
-\frac{1}{4M_*}
\left(
\begin{array}{cc}
A & A \\
-B & B \\
\end{array} 
\right) 
\left(
\begin{array}{cc}
{1 \over \tan[\pi R(\eps -X)]} & 0 \\
0 & {1 \over \tan[\pi R(\eps +X)]} \\
\end{array} 
\right) 
\left(
\begin{array}{cc}
A & -B \\
A &  B \\
\end{array} 
\right) . 
\label{35}
\end{eqnarray}
When $X \simeq \frac{1}{2R}$, 
 even small $\eps$ can
 produce a Majorana mass hierarchy 
 that is large enough 
 for the SRND to work.
In the case of 
 $(\eps + X ) = \frac{1}{2R}$,
 $|\eps - X |$ is also around $\frac{1}{2R}$
 but still 
 effective for the seesaw.
Then 
Eq.(\ref{35}) becomes 
\begin{eqnarray}
m_{\nu}^{\rm 5D}  \simeq  
-{1 \over 4M_* \tan[\pi R(\eps -X)]} 
\left(
\begin{array}{c}
A \\
-\hspace{-0.5mm}B 
\end{array} 
\right) 
(A \;  -\hspace{-1mm}B) ,
\label{224}
\end{eqnarray}
which suggests the NH spectrum. 
Maximal mixing will be  
 obtained in the case of 
 $|A| \simeq |B|$, 
 even if $\eps \ll X$. 

The suitable 3-generation flavor
 structure is obtained
 by introducing a 
 symmetry breaking parameter, $C'$,  
 in the Dirac mass matrix as
\begin{equation}
m = \left(
\begin{array}{ccc}
C &  &  \\
C' & A  & \\
 &    & B
\end{array}
\right),  \;\;\;
M = \left(
\begin{array}{ccc}
Y & & \\
 & \eps& X \\
 & X & \eps
\end{array}
\right) .
\end{equation}
Taking $Y$ 
 to be real and positive 
 as $Y < X$, 
 the 5D seesaw induces 
 the neutrino mass matrix
\begin{equation}
m_\nu^{\rm 5D} 
 \simeq 
-\frac{1}{4M_*}
\left(
\begin{array}{ccc}
\frac{2C^2}{\pi RY} & \frac{2CC'}{\pi RY}  & 0 \\
\frac{2CC'}{\pi RY} & 
\frac{A^2}{\tan[\pi R (\eps-X)]}\hspace{-1mm}
+\hspace{-1mm}\frac{2C'^2}{\pi RY} 
 & 
-\frac{AB}{\tan[\pi R (\eps-X)]} \\
0 &  -\frac{AB}{\tan[\pi R (\eps-X)]}  & 
\frac{B^2}{\tan[\pi R (\eps-X)]}
\end{array}
\right).
\end{equation}
In the case of $|C| \sim |C'|$ and $|A| \simeq |B|$ with
 $\frac{2|C'^2|}{\pi RY} \sim 0.1 \times
  \frac{|A^2|}{\tan[\pi R (X-\eps)]}$,
 the suitable 
 bi-large mixing is obtained.
Although the partial degeneracy
 is needed in the Dirac mass matrix 
 we can 
 realize bi-large mixing with the NH mass spectrum 
 by only two non-large symmetry
 breaking parameters.

\subsection{SRND in $U(1)_F$ flavor symmetry}

In the previous
 sub-section,
 we have shown one example of
 the SRND when the bulk mass
 was $M^V$. 
This sub-section shows 
 some more examples
 with the bulk mass $M^V$ or $M^S$ 
 by introducing 
 a $U(1)_F$ flavor symmetry. 
The following discussion does not change 
 if we take 
 a discrete subgroup of the 
 $U(1)_F$ 
 such as $Z_N$ with 
 enough large $N$ and the same charge assignments. 
We will show that 
 the SRND can work 
 in a much simpler 
 setup in 5 dimensions.

We introduce
 only one scalar bulk field, $\phi$, with
 $U(1)_F$ charge, $-1$.
The zero-mode of $\phi$ takes a real and positive
 constant VEV as
\begin{equation}
\label{eq def eps}
\frac{\langle \phi \rangle}{\sqrt{2\pi RM_*^3}} \equiv \eps\; (< 1).
\end{equation}
Taking the charge assignment of the
 lepton sector as 
\begin{equation}
L_1: 1, \quad  L_2 :0,\quad L_3:0, \quad 
N_1: 2, \quad  N_2 :1,\quad  N_3:0, 
\end{equation}
the Dirac and Majorana mass matrices
 are given by 
\begin{equation}
\label{14}
m \simeq \left( 
\begin{array}{ccc}
\eps ^3 & \eps ^2 & \eps \\
\eps^2 & \eps  & 1\\
\eps^2 & \eps  & 1
\end{array} 
\right) m_0, \;\;\;
M \simeq \left( 
\begin{array}{ccc}
\eps ^4 & \eps ^3 & \eps ^2\\
\eps ^3 & \eps ^2 & \eps\\
\eps ^2 & \eps &1
\end{array} 
\right) M_0,
\end{equation}
respectively.
We take both $m_0$ and $M_0$ 
 to be real and positive for simplicity. 
They represent only the order, and
 each element has ${\cal O}(1)$
 coefficient.
Thus,  
 the ranks of the mass matrices are both 3 in general.
In the diagonal basis of the Majorana mass matrix, 
 Eq.(\ref{14}) becomes 
\begin{equation}
m_R \simeq 
\left( 
\begin{array}{ccc}
 \eps ^3 &  \eps ^2 &  \eps \\
 \eps^2 &   \eps  &   1 \\
 \eps^2 &   \eps  &   1
\end{array} 
\right) m_0 , \;\;\;
M_d  \simeq
\left( 
\begin{array}{ccc}
 \eps^4 &  & \\
 &   \eps^2  &  \\
 &   & 1 
\end{array} 
\right) M_0.
\label{15}
\end{equation}
Then, 
 the seesaw formula 
 is given by 
\begin{eqnarray}
m_\nu 
 &\simeq&
-\frac{m_0^2}{M_0}\left[
\frac{1}{\eps^4}
\left( 
\begin{array}{c}
 \eps^3 \\ 
 \eps^2  \\ 
 \eps^2  
\end{array} 
\right)
(\eps^3 \; \eps^2 \;\eps^2) +
\frac{1}{\eps^2}
\left( 
\begin{array}{c}
\eps^2  \\ 
\eps \\ 
\eps  
\end{array} 
\right)
(\eps^2 \;\eps \; \eps ) 
+
\left( 
\begin{array}{c}
\eps  \\ 
1 \\ 
1 
\end{array} 
\right)
(\eps \; 1  \;1 ) \right].
\label{sd}
\end{eqnarray}
This mass matrix can potentially
 induce bi-large mixing
 with a NH mass spectrum. 
However, 
 the mass spectrum is naively 
 $\eps^2:1:1$ in units
 of $\frac{m_0^2}{M_0}$, 
 and hence, 
 in order to realize the NH spectrum,
 the determinant of the 2-3 
 sub-matrix must be 
 reduced to ${\cal O}(\eps)$ 
 by tuning of
 ${\cal O}(1)$ coefficients. 
The SRND could avoid this
 tuning, but it 
 does not 
 work in this setup because 
 the three terms in Eq.(\ref{sd}) are all
 of the same order.

Now let us examine 
 the 5D case. 
Thanks to the 5D feature, the
 tan(h)-function 
 can change the orders of three terms in Eq.(\ref{sd}). 
At first, 
 we adopt
 $M^S$ for the bulk mass.
In this case the 5D seesaw suggests 
\begin{eqnarray}
m_\nu^{\rm 5D} 
 &\simeq&
-\frac{m_0^2}{2M_*\tanh[\pi RM_0]}\times \nonumber\\
&& \hspace*{-2mm}
\left[
\delta^S_1
\left( 
\begin{array}{c}
 s\eps \\ 
 t  \\ 
 u 
\end{array} 
\right)
(s\eps \;\;\; t \;\; u) +
\delta^S_2
\left( 
\begin{array}{c}
s'\eps  \\ 
t' \\ 
u'  
\end{array} 
\right)
(s'\eps \;\;\;t' \;\; u' ) 
+
\left( 
\begin{array}{c}
a\eps  \\ 
b \\ 
c 
\end{array} 
\right)
(a\eps \;\; b  \;\;c ) \right]\hspace{-1mm}, 
\label{sdms}
\end{eqnarray}
where
 $\delta^S_1 = \frac{\eps^4\tanh[\pi RM_0]}{\tanh[\pi R\eps^4 M_0]}$,
 $\delta^S_2 = \frac{\eps^2\tanh[\pi RM_0]}{\tanh[\pi R\eps^2 M_0]}$,
 and 
 $s,t,\cdots b,c$ are ${\cal O}(1)$ coefficients 
 in which the coefficients of $M_{i}^S$ are
 absorbed. 
$\delta^S_{1,2} \simeq 1$ when
 $\eps^2 M_0 \ll R^{-1}$, while
 $\delta^S_{1}\simeq \eps^4$ and 
 $\delta^S_{2}\simeq \eps^2$ when
 $\eps^4 M_0 \gg R^{-1}$. 
The SRND structure is obtained 
 when $\delta^S_{1,2} \sim 0.1$, in which
 the reduction of the 2-3 sub-matrix determinant  
 is automatic and  
 the mass spectrum becomes 
 $\delta_{1,2}^S\eps^2:\delta_{1,2}^S:1$
 in units of $\frac{m_0^2}{2M_*\tanh[\pi RM_0]}$.

It should be noticed that the
 dominant contribution to the light neutrino
 mass matrix is coming from
 the heaviest Majorana mass, $M_3^S$,
 which is completely  
 different from the ordinal 
 SRND in the 4D setup. 
This is new type
 of the SRND realized by 
 the 5D feature, the tanh-function.
Precisely speaking, 
 this is also 
 possible 
 even in the 
 4D setup in Eq.(\ref{sd}), 
 if $\frac{|\eps^2 M_3|}{|M_2|}\sim \frac{|\eps^4 M_3|}{|M_1|}
 \sim 0.1$ 
 is realized accidentally 
 by the diagonalization
 of Majorana mass matrix $M$ (Eq.(\ref{15})). 
We stress that the 5D setup 
 can 
 naturally realize the SRND structure
 without 
 tuning of 
 ${\cal O}(1)$
 coefficients in the Majorana mass matrix.

For the suitable mixing angles,
 we should choose ${\cal O}(1)$
 coefficients\cite{later}.
For the 2-3 maximal mixing,
 the condition $|b|\simeq |c| (\sim 1)$ is
 needed.
Remembering that 
 bi-large mixing with the NH mass 
 spectrum 
 needs the following 
 four types of sign assignments 
\begin{equation}
\left( 
\begin{array}{ccc}
\eps^2 & \mp\eps & \pm\eps \\
\mp\eps &  1  &   1 \\
\pm\eps &  1  &   1
\end{array} 
\right) , \;\;\;\;\;
\left( 
\begin{array}{ccc}
\eps^2 & \pm\eps & \pm\eps \\
\pm\eps &  1  &  -1 \\
\pm\eps & -1  &   1
\end{array} 
\right) , 
\label{4}
\end{equation}
we can show that 
 the 3rd term in Eq.(\ref{sdms}) alone cannot 
 realize the bi-large mixing. 
Corrections from the 1st and 2nd terms in Eq.(\ref{sdms}) 
 are needed\cite{King:1999cm}
 for the suitable signs 
 of Eq.(\ref{4}). 
For example, 
 to flip the sign of 
 the (2,1), (1,2) elements,
 the relation 
 $|ab| < |\delta^S_1 s t+\delta^S_2 s't'|$ 
 should be satisfied,
 which is impossible 
 when $\delta^S_{1,2} \ll 1$.
However, 
 the experimental data   
 implies that $\delta^S_{1,2}$ is not
 so small, 
 $\delta_{1,2}^S \sim \sqrt{\dms/\dma}$,
 in the NH mass spectrum.
Hence, 
 the above condition
 can be satisfied 
 even by ${\cal O}(1)$ coefficients. 
We should take relatively small (large) 
 values of $a$ ($s,t,s',t'$)
 as the ${\cal O}(1)$ coefficients\cite{later}.
For example, 
 $a=u'=0.5, s=s'=4, b=c=-t=u=-t'=1$ with
 $\eps = 0.2$ and
 $M_0 = 3.3R^{-1}$
 ($\delta_{1,2}^S \simeq 0.1$) 
 induce 
 $\theta_{12}=33^\circ$,
 $\theta_{23}=46^\circ$,
 $\theta_{13}=3.3^\circ$, and 
 $\sqrt{\dms/\dma}
  =0.22$.

Next, let us adopt
 $M^V$ for the bulk mass. 
Remembering that 
 the seesaw scales are 
 strongly split 
 when $|M_i| \sim |\frac{2n+1}{2R}|$,
  we assume 
\begin{eqnarray}
\label{assume}
|M_1| \ll \frac{1}{2R}, \;\;\;
|M_2| \simeq \frac{1}{2R}, \;\;\;
|M_3| \simeq \left|\frac{2n+1}{2R}\right|. 
\label{322}
\end{eqnarray}
The neutrino mass matrix is given by
\begin{eqnarray}
\hspace*{-2mm}
m_\nu^{\rm 5D}\hspace{-2mm} &\simeq & \hspace{-2mm}
-\frac{m_0^2}{2\pi RM_*M_0}\left[
\left( 
\begin{array}{c}
\eps \\ 
1  \\ 
1  
\end{array} 
\right)\hspace{-1mm}
(\eps \:\:\:1 \:\:\:1) + \delta
\left( 
\begin{array}{c}
\eps  \\ 
1 \\ 
1 
\end{array} 
\right)\hspace{-1mm}
(\eps \:\:\: 1 \:\:\:1) 
 + \hat\delta
\left( 
\begin{array}{c}
\eps  \\ 
1 \\ 
1 
\end{array} 
\right)\hspace{-1mm}
(\eps \:\:\: 1 \:\:\:1) 
\right]\hspace{-1mm},
\label{sd1} 
\end{eqnarray}
where $\delta = \frac{\pi R\eps^2M_0}{\tan[\pi R|M_2|]}$ 
 and 
 $\hat\delta = \frac{\pi RM_0}{\tan[\pi R |M_3|]}$. 
Thanks to the 5D feature, the tan-function, 
 the SRND structure can be obtained 
 with $\delta \sim \hat\delta \sim 0.1$.  
This is the similar structure to 
 Eq.(\ref{sdms}), in which 
 the reduction of the 2-3 sub-matrix determinant
 is automatically achieved with 
 the mass spectrum
 $\delta\eps^2:\delta:1$
 in units of $\frac{m_0^2}{2\pi RM_*M_0}$.
${\cal O}(1)$ coefficients
 can induce the suitable mixing angles\cite{later}.

The above model works well but 
 the assumption of Eq.(\ref{assume})
 seems artificial, which
 can be improved by 
 taking 
 new charge assignment
\begin{equation}
L_1: 2, \quad  L_2 :1,\quad L_3:1, \quad
N_1: 2, \quad  N_2 :1,\quad  N_3:-1.
\end{equation}
In this case
 the Dirac and Majorana mass matrices
 become 
\begin{equation}
\label{22}
m \simeq \left(
\begin{array}{ccc}
\eps ^4 & \eps^3 & \eps \\
\eps^3 & \eps^2  & 1\\
\eps^3 & \eps^2  & 1
\end{array}
\right) m_0, \;\;\;
M \simeq \left(
\begin{array}{ccc}
\eps^4 & \eps^3 & \eps\\
\eps^3 & \eps^2 & 1\\
\eps & 1 & \eps^2
\end{array}
\right) M_0,
\end{equation}
respectively.
Each element has an ${\cal O}(1)$
 coefficient and 
 the ranks of mass matrices are 3.
In the diagonal basis of the
 Majorana mass matrix, 
 Eq.(\ref{22}) 
 becomes 
\begin{equation}
m_R \simeq \left( 
\begin{array}{ccc}
\eps^{4} & \eps  & \eps \\
\eps^{3} & 1  & 1\\
\eps^{3} & 1  & 1
\end{array} 
\right) m_0, \;\;\;
M_d \simeq \left( 
\begin{array}{ccc}
\eps^{4} &   &  \\
 & -1\hspace{-1mm}+\hspace{-1mm}\eps^2  & \\
 &    & 1\hspace{-1mm}+\hspace{-1mm}\eps^2 
\end{array} 
\right) M_0.
\end{equation}
Setting 
 $M_0 \simeq \frac{1}{2R}$,
 the 5D seesaw induces 
\begin{eqnarray}
m_\nu^{\rm 5D} &\simeq & 
-\frac{\eps^2 m_0^2}{2 \pi R M_* M_0}\left[
\left( 
\begin{array}{c}
 \eps \\ 
 1  \\ 
 1  
\end{array} 
\right)
(\eps \; 1 \; 1) + \delta'
\left( 
\begin{array}{ccc}
\eps^2 & \eps & \eps  \\ 
\eps & 1 & 1 \\ 
\eps & 1 & 1 
\end{array} 
\right) \right],
\label{sd2} 
\end{eqnarray}
where 
 $\delta' = \frac{\pi R M_0}{\eps^2\tan[\pi R M_0]}$. 
The 2nd 
 mass matrix 
 has rank 2\footnote{If we take 
 $|M_3|=\frac{1}{2R}(={\cal O}((1+\eps^2)M_0))$,
 the 2nd mass matrix 
 in Eq.(\ref{sd2})
 is only produced by 
 $|M_2|={\cal O}((-1+\eps^2)M_0)$, and has rank 1.}.
The 
 coefficients of  
 $M_{i}$ are absorbed into 
 those of the mass matrices in 
 Eq.(\ref{sd2}). 
The SRND structure is obtained when $\delta' \sim 0.1$, 
 where
 the 1st term in Eq.(\ref{sd2})
 is dominant.
The reduction of the determinant 
 in the 2-3 sub-matrix 
 is automatic and 
 the NH mass spectrum becomes 
 $\delta'\eps^2 :\delta':1$
 in units of 
 $\frac{\eps^2 m_0^2}{2 \pi R M_* M_0}$. 
Again, ${\cal O}(1)$ coefficients
 can induce the suitable mixing angles\cite{later}.

\vspace{2mm}

Finally, let us show 
 one more example of the application of 
 the tanh-function. 
We take the
 charge assignment to be 
\begin{equation}
L_1: 1, \quad  L_2 :-1,\quad L_3:-1, \quad
N_1: 1, \quad  N_2 :0,\quad  N_3:0.
\end{equation}
Then, the Dirac and Majorana mass matrices
 are given by
\begin{equation}
m \simeq \left(
\begin{array}{ccc}
\eps^2 & \eps & \eps \\
 1 & \eps  & \eps \\
 1 & \eps  & \eps
\end{array}
\right) m_0, \;\;\;
M \simeq \left(
\begin{array}{ccc}
\eps ^2 & \eps    & \eps  \\
\eps & 1 & 1\\
\eps & 1 & 1
\end{array}
\right) M_0,
\end{equation}
respectively.
The Majorana mass spectrum is 
 $\eps^2: 1: 1$ in units of $M_0$. 
In the diagonal basis of the 
 Majorana mass matrix, 
 the Dirac mass matrix
 does not change the order
 of each element. 
Then, the light neutrino mass matrix is
 given by
\begin{eqnarray}
m_\nu^{\rm 5D} &\simeq & 
-\frac{m_0^2}{2 M_*\tanh [\pi R \eps^2 M_0]}\left[
\left( 
\begin{array}{c}
 \eps^2 \\ 
 1  \\ 
 1  
\end{array} 
\right)
(\eps^2 \;\;\; 1\;\;\; 1) + \delta''
\left( 
\begin{array}{ccc}
\eps^2 & \eps^2 & \eps^2 \\ 
\eps^2 & \eps^2 & \eps^2 \\ 
\eps^2 & \eps^2 & \eps^2 
\end{array} 
\right) \right],
\label{sd3} 
\end{eqnarray}
where
 $\delta'' = \frac{\tanh[\pi R \eps^2 M_0]}{\tanh[\pi R M_0]}$.
The 2nd mass matrix has rank 2.
The seesaw 
 scales become degenerate 
 due to the tanh-function 
 when $|M_1^S| > {\cal O}(R^{-1})$. 
Both $\tanh[\pi R \eps^2 M_0]$ and $\tanh[\pi RM_0]$
 become 1 ($\delta''\simeq1$),
 and hence
 Eq.(\ref{sd3}) becomes
\begin{eqnarray}
m_\nu^{\rm 5D} &\simeq & 
-\frac{m_0^2}{2 M_*}
\left( 
\begin{array}{ccc}
\eps^2 & \eps^2 & \eps^2 \\ 
\eps^2 & 1 & 1 \\ 
\eps^2 & 1 & 1
\end{array} 
\right). 
\end{eqnarray}
Notice that 
 the determinant of the 
 2-3 sub-matrix is automatically
 reduced to ${\cal O}(\eps^2)$.  
This mass matrix can 
 induce bi-large mixing
 with a NH mass spectrum by suitable 
 ${\cal O}(1)$ coefficients\cite{later}.


\section{Summary}

In the 5D theory in which the 
 3 generation right-handed neutrinos
 are in the bulk, 
 the neutrino flavor mixings and the mass spectrum
 can be constructed through the 
 5D seesaw.
We have presented 
 a simple calculation method 
 of the 5D seesaw, 
 in which one just 
 replaces the Majorana mass eigenvalues
 according to Eqs.(\ref{seesaw5D3}) and (\ref{11}). 
This technique 
 should be useful for studies of 
 neutrino mass textures. 
The neutrino flavor mixings and the mass spectrum
 can be drastically 
 changed by the 
 5D seesaw.
The 5D 
 features
 appear 
 when
 the Majorana mass is the same
 as the compactification scale 
 or
 larger than it.  
Depending on the 
 type of the bulk mass, 
 the seesaw scales of the 3 generations 
 are strongly split or 
 degenerate.
In the split case, 
 the seesaw enhancement
 is naturally realized. 
We have shown that   
 the SRND 
 works in a simple
 setup in 5 dimensions. 
This enhancement of the effective Majorana mass 
 hierarchy
 makes 
 the tiny symmetry breaking 
 induce the suitable flavor structure
 in the $L_\mu-L_\tau$
 symmetry. 
The 5D seesaw 
 can naturally derive 
 some specific mass textures 
 which are just assumptions in 
 the 4D setup.
The degenerate case  
 can also 
 realize the suitable 
 flavor structure.

We have only considered 
 the situation that all the 3 generations' 
 seesaw scales are characterized 
 either by the tan- or tanh-function.
It should be considerable that 
 different generations have 
 different seesaw scales characterized by
 tan- or tanh-functions 
 (Eqs.(\ref{mkms}) and (\ref{msmk})) and 
 induce the neutrino flavor structure.
The intermediate scale
 (such as $10^{9 \sim 13}$ GeV) can be 
 effectively obtained even by 
 lower scales of $M_*, M^V$, $R^{-1}$ due to 
 the tan-function. 
It is also valuable to 
 examine the 
 leptogenesis in this framework. 
The extension to the warped extra
 dimension\cite{Randall:1999ee} might also induce 
 new possibilities\cite{Huber:2003sf}.

\vskip 1cm

\leftline{\bf Acknowledgments}
N.H. is supported by the Alexander von Humboldt Foundation,
 and would like to thank 
 M.-T. Eisele, M. Lindner, and K. Yoshioka 
 for fruitful discussions and
 collaboration in part. 
N.H. would also like to
 thank 
 W. Rodejohann, M. Schmidt,
 A. Merle, E. Akhmedov,  
 T. Ota, N. Okada, 
 and T. Yamashita 
 for useful discussions.

\vspace{.5cm}


\end{document}